\input FEYNMAN
\documentclass[amsmath]{amsart}
\usepackage{graphicx}

\usepackage{amsmath}
\usepackage{amsfonts,amssymb}
\usepackage{latexsym}
\usepackage{rsfs}
\usepackage{latexsym}
\usepackage{rsfs}

\begin{document}

%
%

\title{Planar Homological Mirror Symmetry
}

\author{Eiji Konishi}

\address{Department of Physics, School of Science, Kyoto University,
 Sakyo-ku, Kyoto 606-8502, Japan
 }
\email{
Konishi.Eiji@s04.mbox.media.kyoto-u.ac.jp}
\date{\today}
\maketitle
%
\date{\today}
\newcounter{ichi}
\setcounter{ichi}{1}
\newcounter{ni}
\setcounter{ni}{2}
\newcounter{san}
\setcounter{san}{3}
\renewcommand{\theenumi}{\roman{enumi}}
\newcommand{\cqfd}{\hfill $\square$}
\begin{abstract}In this article, we formulate a planar limited version of the B-side in homological mirror symmetry that formularizes Chern-Simons-type topological open string field theory using homotopy associative algebra ($A_{\infty}$ algebra). This formulation is based on the works by Dijkgraaf and Vafa. We show that our formularization includes gravity/gauge theory correspondence which originates in the AdS/CFT duality of Dijkgraaf-Vafa theory.
\end{abstract}

\section{Introduction}
In a certain large $N$ limit of $SU(N)$ gauge theory, it is believed that it can be described by closed strings. Under this limit of the open string world sheet, termed the planar limit, many physical theories originating from different fields organize and cross each other.

In this article, we combine the fruitful example of large $N$ duality, Dijkgraaf-Vafa duality, that exists between topological {B}-model string fields as holomorphic Chern-Simons fields and integrable systems in Seiberg-Witten theory as sectors of a one matrix model as detailed in Refs. 1-6 with homological mirror symmetry conjecture between the topological {A}-model and {B}-model. \cite{D,E} 

In the Dijkgraaf-Vafa conjecture, the integrable system of Seiberg-Witten theory is determined from one matrix model, for example, the prepotential is determined from the matrix free energy, and the Seiberg-Witten curve and differential are obtained from these values of the matrix model as explained in sec.3.1.\cite{C,C1,C2}

 We derive a string duality covariant equation of topological open string field theory from these dualities.

The proposed equation has not been proved mathematically. We mathematically conjecture the statements described in this article as follows.

First, we summarize this article's formulation of the Dijkgraaf-Vafa conjecture.
The Dijkgraaf-Vafa conjecture of the {B}-model gives the $A_{\infty}$ categorical injective functor $\mathfrak{DV}$ from the $A_{\infty}$-enhanced derived special geometrical $A_{\infty}$  category for a $SU(\infty)$ Seiberg-Witten curve $C$ and prepotential $\mathcal{F}$ to the $A_{\infty}$-enhanced derived category of the {B}-model:
\begin{equation}\mathfrak{DV}:\mathcal{D}_{\infty}\mathcal{F}(C)\to\mathcal{D}_{\infty}^b(Coh(V^{\dagger}))\end{equation}
where $V^{\dagger}$ is a Calabi-Yau threefold specified as the target space
and $\mathcal{F}(C)$ is $A_{\infty}$ category defined in special geometry of the matrix model which shall be defined in sec.2.3.

If the above conjecture holds true, we can interpret {B}-branes, which are associated with holomorphic bundles on the Calabi-Yau manifold $V^{\dagger}$, as an infinite-dimensional integrable structure of KP/Toda-solitons.

As one of the results detailed in this article, the statement of mirror symmetry is summarized by the following formulation.

\smallskip
\smallskip
\textit{The equivalence of product structure in planar limited homological mirror symmetry conjecture for the {A}-model target Calabi-Yau threefold $V$ under the mirror {B}-model Dijkgraaf-Vafa conditions becomes the following equation described using reduced $A_{\infty}$ algebras with matrices $\Phi$ and $\Phi^{\dagger}$:
\begin{equation}\mathfrak{F}(\lambda\,|\>\Phi^{\dagger})\simeq \tau(\mathcal{T}\!\cdot\Phi)\end{equation}
where disk amplitude $\mathfrak{F}$ is the $A_{\infty}$ algebra of a special function in the general product structure of Floer cohomology in the mirror Calabi-Yau threefold $V$.}
\smallskip
\smallskip

Regarding the above conjecture, the statement is not yet complete, since this article focuses on the right-hand side, and does {not} treat the mathematical modification of the left-hand side, i.e, Floer cohomologies.
This term $\mathfrak{F}$ is Fukaya's generalization of his multi-theta function $\vartheta$. \cite{F,F1} 
Terms that appear in the above conjectures are defined in subsequent sections.
\section{$A_{\infty}$ Form in Topological Quantum Field Theory}
\subsection{Mathematical definition of $A_{\infty}$ category}
The covariant equation is based on the $A_{\infty}$ categorical construction for topological quantum field theory.

First, we define the $A_{\infty}\!$ category. \cite{G}

The objects of the $A_{\infty}$ category $\mathcal{C}$ are elements of a set $Ob(\mathcal{C})$. 
For two elements of this set $c_1,c_2\in Ob(\mathcal{C})$, the (co)chain complex $\mathcal{C}_*(c_1,c_2)$ is defined.

 For the natural numbers $k=1,2,\ldots$, the linear map termed the \textit{cup product} and \textit{Massey product}
\begin{equation}\eta_k:\mathcal{C}_*(c_0,c_1)\otimes\cdots\otimes \mathcal{C}_*(c_{k-1},c_k)\to \mathcal{C}_*(c_0,c_k)\end{equation}
is defined for each object $c_i\in Ob(\mathcal{C})$ and elements $x_i\in \mathcal{C}_*(c_i,c_{i+1})$. 

These satisfy the following $A_{\infty}$ conditions.
\begin{itemize}
\item $\eta_2$ is a (co)chain map.
\item  For ${\mathrm{sgn}_i=(-)^{\sum_{k=1}^{i-1}(\deg x_k+1)}}$ and $\partial\eta_3=\partial\cdot\eta_3+\mathrm{sgn}\cdot\eta_3\cdot\partial$, the following relation stands.
\begin{equation}(\partial\eta_3)(x_1\otimes x_2\otimes x_3)=\mathrm{sgn}_{1}\cdot \eta_2(\eta_2(x_1\otimes x_2)\otimes x_3)+\mathrm{sgn}_2\cdot \eta_2(x_1\otimes \eta_2 (x_2\otimes x_3)).\end{equation}
\item In addition, the following relation stands.
\begin{eqnarray}&&(\partial\eta_k)(x_1\otimes\cdots\otimes x_k)\nonumber\\&&=\sum_{1\le i\le j\le k}\mathrm{sgn}_i\cdot\eta_{k-j+i}(x_1\otimes \cdots\otimes\eta_{j-i}(x_i\otimes \cdots\otimes x_j)\otimes\cdots\otimes x_k).\end{eqnarray}
\end{itemize}

This is the mathematical definition of the $A_{\infty}$ category.
We note that the higher composition structure is a matrix $\eta_{ij}^k$, having two indices concerning two objects $1\le i,\, j\le k$, whose elements are Massey product functions $\eta_l,$ $2\le l\le k$.
\subsection{$A_{\infty}$ category of topological open string {A}-model}
\subsubsection{Fukaya category $Fuk(V)$}
The $A_{\infty}$ category of the topological {A}-model $Fuk(V)$, termed the \textit{Fukaya category}, is defined as follows. 

Objects of the Fukaya category are {A}-branes, i.e., pairs of Lagrange submanifolds $\mathcal{L}$ of a Calabi-Yau threefold $V$, namely the half-dimensional submanifold $\mathcal{L}$ where the symplectic two-form $\omega$ of $V$ vanishes 
\begin{equation}\dim_{\mathbf{R}}\mathcal{L}=3,\ \omega|_{\mathcal{L}}=0,\end{equation}
and Chan-Paton bundles $\mathcal{E}$ on $\mathcal{L}$. 
 The cochain complex $\mathcal{C}^V_*(\mathcal{L}_{*_1},\mathcal{L}_{*_2})$ is defined by one of the relative Floer cohomologies of two Lagrangians and Chan-Paton bundles, namely BRST cohomology concerning BRST charge $Q$: 
\begin{equation}\mathcal{C}_k^V(\mathcal{L}_i,\mathcal{L}_j):=H^k((\bigoplus_{\mathcal{L}_i\cap \mathcal{L}_j,\eta(p)=k}{C}[p],Q);{C})\end{equation}
where $\eta$ is the ghost number of open string instanton, known as the Maslov index, and $(\bigoplus_{\mathcal{L}_i\cap \mathcal{L}_j,\eta(p)=k}$ ${C}[p],Q)$ is the Morse-Witten complex.\cite{G0}

 The simplified Massey product $\eta_k$ for $[q_{i-1,i}]\in \mathcal{C}^V_*(\mathcal{L}_{i-1},\mathcal{L}_{i})$ is defined in the Ref. 7 by
\begin{equation}
\eta_k([q_{0,1}]\otimes\cdots\otimes [q_{k-1,k}])=\sum_{ [q_{k,0}]
}\sum_{\mathrm{ord}(\partial D^2)}\sum_{\phi}\mathrm{sgn}([q_{k,0}])\cdot\exp(-\int_{D^2}\phi^*\omega)[q_{k,0}]
\end{equation}
where holomorphic maps $\phi:D^2\to V$ are homotopy classes ${\phi\in\pi_2(V,\bigcup_{i=1}^{k+1} \mathcal{L}_i)}$, i.e., embedding maps of the world sheet disk $D^2$ to target space $V$, and each map $\phi$ preserves the cyclic orders of marked boundaries $\mathrm{ord}(\partial D^2)$ on the disk. 

\begin{figure}[ht]
\scalebox{.3}{\includegraphics{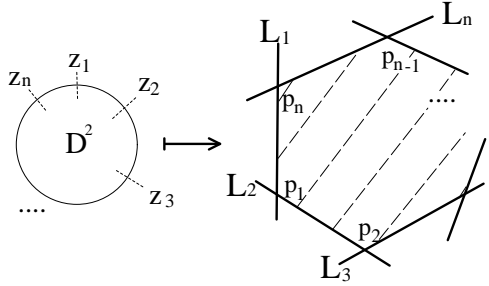}}
\caption{Embedding map $\phi$ from disk $D^2$ into target space $V$}
\end{figure}

In figure 1, for $n$ number of coordinates of asymptotic states $z_i\in\partial D^2$, $\phi(z_i)=p_i\in\mathcal{L}_i\cap\mathcal{L}_{i+1}$ stands.
Generally, the coefficient function $\sum_{\mathrm{ord}(\partial D^2)}\sum_{\phi}
\pm$ $\exp(-\int_{D^2}\phi^*\omega)$ is modified to a special function $\mathfrak{F}$. 
 In sec.2.2.2, we give a concrete example of this special function.
\subsubsection{Example of Massey product of Fukaya category for tori}
In this subsection, in accordance with Ref. 13 we define a physically modified product structure appearing in a Fukaya multi theta function $\vartheta$, i.e., the product structure of the Fukaya category for tori as the simplest model in general special functions.

 The function $\vartheta$ is defined by the generating function for counting the number of $0$-loop world sheets with boundaries on the {A}-brane. \cite{H} 

The product structure of Floer cohomologies in the torus $V$ is expressed by a matrix of Fukaya multi theta functions.
\begin{equation}\eta_k:\mathcal{C}^V_{*}(\mathcal{L}_0,\mathcal{L}_1)\otimes\cdots\otimes \mathcal{C}^V_*(\mathcal{L}_{k-1},\mathcal{L}_{k}) \to \mathcal{C}^V_{*}(\mathcal{L}_0,\mathcal{L}_k)\end{equation}
 where flat torus $\mathcal{L}_i$ is parameterized by configuration in $V$ denoted by $\nu_i$.

We let $\vec{\mathscr{A}_i}$ be composed of flat connections on Lagrangians, i.e., gauge potentials as Chan-Paton factors whose curvature is given by the modified symplectic form presence of the B-field, $F=\omega+\sqrt{-1}B$, and let $B$ be the B-field two-form combined with open strings, and $\vec{\mathcal{L}}$ denote the set of $k+1$ Lagrangians.

For an embedding map $\phi:D^2\to V$, we write the area of the world sheet as \begin{equation}S_k(\vec{\nu})=\int_{D^2}\phi^*F.\end{equation}

We set counting of the number of $0$-loop world sheets with $k+1$ boundaries between $k+1$ {A}-branes as $\mathscr{C}_k(\vec{\mathcal{L}})$.

The Fukaya multi theta function with coefficients $\mathscr{C}_k$ is defined as \cite{H}
\begin{equation}
\vartheta_k(\vec{\nu}\,|\>\vec{\mathscr{A}};F)=\sum_{\vec{\gamma}} \mathscr{C}_k(\vec{\nu}+\vec{\gamma})\exp(-2\pi(S_k(\vec{\nu}+\vec{\gamma})+\sqrt{-1}\sum_i\mathscr{A}_i\Delta_{i-1,i}(\phi(z_{i}))))
\end{equation}
where the sum is taken over all 
the configurations of Lagrangians $\mathcal{L}(\vec{\nu}+\vec{\gamma})$.
The right-hand side of eq.(11) stands for general Calabi-Yau target spaces.

The Massey product $\eta_k$ is defined for $[q_{i,j}]\in \mathcal{C}^V_*(\mathcal{L}_i,\mathcal{L}_j)$ by
\begin{equation}
\eta_k([q_{0,1}]\otimes\cdots\otimes[q_{k-1,k}])=\sum_{[q_{k,0}]}\vartheta_k(\vec{\nu}\,|\>\vec{\mathscr{A}};F)|_{([q_{0,1}],\ \cdots,\ [q_{k,0}])}[q_{0,k}].
\end{equation} 

To derive the first $A_{\infty}$ relation eq.(4), we adopt a Batalin-Vilkovisky master equation for $\mathscr{C}_k^{(l)}$:
\begin{equation}d\mathscr{C}_k^{(l)}+\sum_{l_1+l_2=l+1}\sum_{k_1+k_2=k+1}\mathrm{sgn}_{l_1,l_2,k_1,k_2}\cdot \mathscr{C}_{k_1}^{(l_1)}\cdot \mathscr{C}_{k_2}^{(l_2)}=0
\end{equation}
where $\mathscr{C}_k^{(l)}$ is the generalization of $\mathscr{C}_k=\mathscr{C}_k^{(0)}$ to general degree $l$ current.

The second $A_{\infty}$ relation eq.(5) is also derived.

These $A_{\infty}$ relations eq.(4) and eq.(5) on $Fuk(V)$ correspond to the perturbative expansion of the generating partition function $Z_{\vartheta}$ of $\vartheta$ functions.

\subsection{$A_{\infty}$ category of Seiberg-Witten geometry}
\subsubsection{Brief review of Seiberg-Witten geometry}
We recall that Seiberg-Witten theory consists of the following three pieces of data.

\begin{itemize}
\item  A Riemann surface, in particular, an (hyper)-elliptic curve $C$ termed a \textit{Seiberg-Witten curve}.

\item A family of Seiberg-Witten curves $\mathcal{M}_{SW}$.

\item A meromorphic one-form $dS^g_{SW}$ on $C$ for genus $g$ of $C$.
\end{itemize}

These pieces of data determine the monodromy property related to vacuum transitions in $\mathcal{N}=2$ supersymmetric Yang-Mills theory.

In the homology cycles of curve $C$, $A_i, B_i \in H_1(C,{Z})$, electric $a_i$ and magnetic $a_i^D$ periods are defined as \begin{equation}a_i=\oint_{A_i}dS^g_{SW},\quad a_i^D=\frac{\partial \mathcal{F}}{\partial a_i}=\oint_{B_i}dS^g_{SW}.\end{equation}

The integrable hierarchies behind Seiberg-Witten data is revealed from the equation $\partial dz_j/\partial a_i=\partial dz_i/\partial a_j$ $(1\le i,j\le N-1)$ for normalized holomorphic differential $\oint_{A_i}dz_j=\delta_{ij}$ that is obtained from derivation of Seiberg-Witten differential by moduli parameters with certain conditions. These conditions are compatibility conditions of an integrable system for the solution thereof $dS$. This is the starting point.
The $SU(N)$ Seiberg-Witten curve can be understood as being the spectral curve of the $N$-periodic Toda chain. \cite{J,K,K1,K2,IM1,IM2}

We construct the KP/Toda $\tau$ function and prove it to be the product structure of the homotopical category $\mathcal{F}(C)$ in two steps a and b.

a. Whitham-deformed Seiberg-Witten differential $d\tilde{S}^g$.

The solution of $N$-periodic Toda chain system is a wave function concerning the labeling $n$ and time variable $t$ of this system and additional parameters $\vec{\theta}$. These parameters correspond to slow variables in the Whitham hierarchy $T_0,$ $T_1$ and $\vec{a}$, respectively. \cite{K,K1,K2}

On the basis of the above facts, we define the Whitham deformation of the original Seiberg-Witten data more precisely.

The Whitham deformation is the replacement of fast time $t_i$ in an integrable system with slow time $T_i$ as $T_i=\epsilon
(t_i)$, for constant $\epsilon<<1$.

The Whitham-deformed Seiberg-Witten differential $d\tilde{S}^g$ is a linear combination of meromorphic differentials and normalized holomorphic differentials by period $a_i$ and restricted slow time variables $T_n$ and $\bar{T}_m$:
\begin{equation}d\tilde{S}^g=\sum_{i=1}^ga_id\omega_i+\sum_{n:odd}T_{n}d\Omega_{n}+\sum_{m:odd}\bar{T}_{m}d\bar{\Omega}_{m}\end{equation} 
where the differential $d\Omega_n$ is determined using the following two conditions.
\begin{itemize}
\item In the limit of $z\to\infty$, $d\Omega_n$ locally becomes
\begin{equation}
d\Omega_{2n+1}=dz^{(2n+1)/2}+holomorphic\ term.
\end{equation}
\item The integration of $\Omega_{2n+1}$ along the $A_i$-cycle vanishes:
\begin{equation}
\oint_{A_i}d\Omega_{2n+1}=0.
\end{equation}
\end{itemize}

The Whitham-deformed Seiberg-Witten differential $d\tilde{S}^g$ reduces to the Seiberg-Witten differential $dS^g_{SW}$ for slow time variables $\vec{T}=(1,0,0,\ldots)$.

  The Whitham deformation satisfies the Whitham equation for time variables. As mentioned in Ref. 20, this equation is written for a prepotential $\mathcal{F}^g(\vec{a},\vec{T})$ as\begin{equation}\frac{\partial^3\mathcal{F}^g(\vec{a},\vec{T})}{\partial T_n\partial a^2}=s_n(\mathcal{F}^g(\vec{a},\vec{T}))\frac{\partial^3 \mathcal{F}^g(\vec{a},\vec{T})}{\partial a^3}\end{equation}
 for functionals $s_n$, $n:odd$.

b. The KP/Toda $\tau$ function corresponding to $SU(g+1)$ Seiberg-Witten theory $\tau_g(\vec{t})$.

First, we note that 
these Seiberg-Witten data are obtained from an integrable system for the case of $SU(2)$ from the KdV equation and for the case of $SU(g+1)$ from the $g$-period Toda equation. \cite{J}

Next, we define the KP/Toda $\tau$ function. 

The Whitham $\tau$ function of the KP/Toda hierarchy $\tau_g(\vec{a},\vec{T},\vec{V}_n)$ is defined by the Toda lattice time variable hierarchies $\vec{T}=(T_1,T_3,\ldots)$ and $\vec{t}=(t_1,t_3,\ldots)$ as 
\begin{equation}
\tau_g(\vec{a},\vec{T},\vec{V}_n)=\exp(2\pi i\mathcal{F}^g(\vec{a},\vec{T}))\,\vartheta_g(\sum_{n:odd} \epsilon^{-1}_n(T_n)\cdot\vec{V}_n|T)|_{\vec{T}=(1,0,0,\cdots)}.
\end{equation}
Here, ${\vec{V}_n=V_j^{(n)}=\frac{1}{2\pi i}\oint_{B_j}d\Omega_n}$ is a $g$-dimensional vector and ${T_{jk}=\frac{1}{2\pi i}\oint_{B_j}d\omega_k}$ is a $g\times g$-period matrix for genus $g$ of the Seiberg-Witten curve $C$. This matrix specifies the complex moduli of $C$.

 The KP/Toda $\tau$ function is originally a function of time variables $\vec{t}$, but in this case it becomes a function of the Seiberg-Witten moduli $\vec{a}$ and Toda time variables $\vec{T}$.
This function ${\vartheta_g(\sum_{n:odd} \epsilon^{-1}_n(T_n)\cdot\vec{V}_n|T)}$ is a $g$-dimensional Riemann theta function and has $Sp(2g,{Z})$ modularity. 

 The classical prepotential $\mathcal{F}_{cl}^g(\vec{a},\vec{T})$ is given by the periods of the deformed Seiberg-Witten differential $d\tilde{S}^g$ as 
\begin{equation}
\mathcal{F}^g_{cl}(\vec{a},\vec{T})=\frac{1}{2}\sum_{i=1}^g\oint_{A_i}d\tilde{S}^g\oint_{B_i}d\tilde{S}^g:=\frac{1}{2}\sum_{i=1}^g\mathcal{F}_i^g(\vec{a},\vec{T}),
\end{equation}
where $\mathcal{F}_i$ are defined for an independent sum over homological cycles $\mathcal{A}_i:=\sum_j \alpha^j_iA_j,$ $\alpha_i^j$$\in{Z}$.
Thus, the classical prepotential $\mathcal{F}_{cl}^g$ becomes
\begin{equation}
\mathcal{F}_{cl}^g=\frac{1}{2}\sum_{i,j=1}^ga_ia_j\tau^{cl}_{ij}+\frac{1}{2}\sum_{n:odd}T_n(\sum_{i=1}^ga_iV_i^{(n)}).
\end{equation}
This formula consists of two terms, i.e., the quadric form of the Seiberg-Witten moduli with (effective) coupling constants $\tau_{ij}^{cl}$ and a combination of the Toda lattice time variable and Seiberg-Witten moduli. 
The logarithm of the KP/Toda $\tau$ function reduces to the Seiberg-Witten prepotential ignoring orders of the slowing parameter of less than more than minus one $\epsilon^n$ $(n\ge -1)$ terms.\cite{K,K1,K2}
 
In the next definition of the $A_{\infty}$ category, we need the higher genus compact Riemann surface $\Sigma^1=\Sigma_g\sharp\Sigma_h$ under the limit $h\to\infty$. We define the $g$-th composition structure $\tau_g$ on the surface $\Sigma^2=\Sigma_g\sharp S^2$, under the conditions
\begin{equation}
\oint_{A_i^1}dS_{SW}^{g+h}=\oint_{A_i^2}dS_{SW}^g\ and\ \oint_{B_j^1}dS_{SW}^{g+h}=\oint_{B_j^2}dS_{SW}^g,
\end{equation}
for all symplectic bases on these surfaces, $A_i^{1\ (2)},B_j^{1\ (2)}\in H_1(\Sigma^{1\ (2)},{Z})$.

\subsubsection{$A_{\infty}$ category $\mathcal{F}(C)$}

First, we explain the evidence for existence of the $A_{\infty}$ category of the prepotential of Seiberg-Witten data.

The $A_{\infty}$ category $Fuk(V)$ is able to be defined because of the existence of the Massey product. This product exists due to the fact that the Feynman diagram of an open string is two-dimensional disk $D^2$ in all perturbative degrees. 

In point particle field theory, a Feynman diagram describes a one-dimensional path. Therefore, one is not able to make a Massey product in the same way. To make a one-dimensional $A_{\infty}$ category, we consider some flows such as renormalization group flows, which are defined on the $(t,g)$ plane by the equation $\frac{\partial \bar{g}(t,g)}{\partial t}=\beta(g)\frac{\partial \bar{g}(t,g)}{\partial g}$ for coupling constant $g_s$, beta function $\beta$ and renormalization time variable $t$, instead of Feynman diagrams. In this strategy, the time variables that determine flows correspond to a coherent state of open string fields.

Next, we explain the $A_{\infty}$ category of Seiberg-Witten theory under the prepotential $\mathcal{F}$. In this case, two facts enable definition of the $A_{\infty}$ category: one is that flows termed \textit{Whitham flows} exist, and the other is that for the prepotential $\mathcal{F}$, the WDVV equation stands;\cite{L,L1,L2,L3,L4,L5} therefore, the s,t-channel duality of Mandelstam variables also stands and we can obtain associative higher compositions.
 
 We use the complex moduli parameters of a holomorphic curve with genus $g$ as a $g\times g$ matrix $T_{ij}$. In the case of a Seiberg-Witten curve, it is shown that for the prepotential $\mathcal{F}$ and Seiberg-Witten moduli $a_i$, $T_{ij}=\frac{\partial^2\mathcal{F}}{\partial a_i\partial a_j}$ stands. These terms are fixed to define a Whitham flow on the moduli space $\mathcal{M}_{dSW}$. 

 These facts enable us to consider flows classified using associative diagrams such as open string world sheets. We shall define the $A_{\infty}$ category in Seiberg-Witten theory termed $\mathcal{F}(C)$ for a Seiberg-Witten curve $C$.
 
In the definition of the prepotential $A_{\infty}$ category, as objects, we take $SU(\infty)$ Seiberg-Witten moduli parameters associated with one of the symplectic bases $A_i$ and $B_i$ of $C$ and denote these by $c_i$. These can be interpretated as BPS monopoles.

In the definition of the morphism, we consider the family of Whitham flows with the prepotential $\mathcal{F}^g$ denoted by $W\!\!f(\mathcal{F}^g)$ on the moduli space $\mathcal{M}_{dSW}$.

 We define the set of morphisms between $c_i$ and $c_j$ as the formal {C} module:
\begin{equation}
\mathcal{C}^C(c_i,c_j):=
\left\lbrace
\begin{array}{c|cc}
 & \mathcal{F}^g\ satisfies\ PDE\ (18)\ for\ a=a_*\ \!and\ \!T=T_*& \\
\, [\mathcal{F}^g(\vec{a},\vec{T})] &T_{*_1*_2}=\frac{\partial^2\mathcal{F}^g}{\partial a_{*_1}\partial a_{*_2}},\!\ V_{*_1}^{*_2}\!=\frac{\partial^2\mathcal{F}^g}{\partial a_{*_1}\partial T_{*_2}}&\\
& \mathcal{F}^g(\vec{a},\vec{T}_{in})=\mathcal{F}_i(\vec{a},\vec{T}_{in}),\ \mathcal{F}^g(\vec{a},\vec{T}_f)=\mathcal{F}_j(\vec{a},\vec{T}_f)&
\end{array}\right\rbrace\end{equation}
for initial and final time variables $\vec{T}_{in}$ and $\vec{T}_f$,
where we define the morphisms as classes of Whitham flows $[\mathcal{F}^g(\vec{a},\vec{T})]$ in $W\!\!f(\mathcal{F}^g)$ as the \textit{Virasoro-constrained in-coming open string states} $|V^{\otimes n}\rangle$ in the S-matrix representation of $\tau$ function $\exp(2\pi i\mathcal{F})=(\bigotimes_{i}\langle \vec{T}_i|)|V^{\otimes n}\rangle$ on corresponding {B}-model target space, i.e., resolved conifold. Then, this space of morphism has the natural structure of open string Hilbert space.
We define the boundaries of $[\mathcal{F}_{ij}]$ as ${\partial_{+\ (-)}[\mathcal{F}_{ij}]:=\lim_{\vec{T}\to\vec{T}_{in\ (f)}^{ij}}[\mathcal{F}_{ij}]}$ and the closure as $\bar{[\mathcal{F}]}=[\mathcal{F}]\cup\partial_+[\mathcal{F}]\cup\partial_-[\mathcal{F}]$. 

The main statement of this article is as follows.

\smallskip
\smallskip
\textit{In the special geometrical homotopical category $\mathcal{F}(C)$, the $k$-th higher composition structure is a matrix whose elements are the linear combination of the KP/Toda $\tau$ function restricted to the sub-Lie group $SU(k)\subset SU(n)$ and substituted Toda time variables.}

 \subsubsection{Completion of main statement}
Here, we complete the main statement. 
We define the $l$-th module of the (co)chain complex $(\mathcal{C}_*^C,\delta)$ as \begin{equation}\mathcal{C}^C_l(c_i,c_j):=\mathcal{C}^C(c_i,c_j)|_{\mathrm{rank}=l}.\end{equation}
The class $[\mathcal{F}_{ik}^{p,\ l_3}]$ is obtained from the deformation such that firstly we deform the time variables in $\bar{[\mathcal{F}_{ij}^{l_1}]}\cup\bar{[\mathcal{F}_{jk}^{l_2}]}$ as $\vec{T}^{ij}_{in}\to \vec{T}^{ij}_m$, $\vec{T}^{jk}_{in}\to \vec{T}^{jk}_m$ and we define \textit{vertex} $V$ and medium time variable $\vec{T}_m$ as $V_{ijk}:=\mathcal{F}_{ij}(\vec{T}_m)=\mathcal{F}_{jk}(\vec{T}_m)$. Secondly we combine them at each vertex, and finally we deform the moduli parameters continuously as $\vec{T}_m\to \vec{T}^{ij}_f$ and $\vec{T}_m\to \vec{T}^{jk}_{f}$.
We put ${\partial[{\mathcal{F}^{ij}_{cl}}]:=\bigcup_{V\in[\bar{\mathcal{F}}_{ij}]}V}$.
 
This definition is based on the Dijkgraaf-Vafa duality between topological {B}-model and $SU(N+1)$ Seiberg-Witten theory. In the definition we take the limit of $N\to\infty$.

We next define the $*$ product, i.e., the composition of two morphisms.

\begin{figure}[ph]
\begin{center}
\begin{picture}(10000,5000)(0,0)
\drawline\fermion[\E\REG](3000,2000)[5000]
\drawarrow[\E\ATTIP](\pmidx,\pmidy)
\drawline\fermion[\NW\REG](\particlefrontx,\particlefronty)[4000]
\drawarrow[\SE\ATTIP](\pmidx,\pmidy)
\drawline\fermion[\SW\REG](\particlefrontx,\particlefronty)[4000]
\drawarrow[\NE\ATTIP](\pmidx,\pmidy)
\end{picture}
\end{center}
\caption{Schematic drawing of the vertex between two Whitham flow classes}
\end{figure}
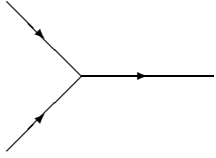

  The rank $l$ of a class $[\mathcal{F}_{*_1*_2}]$ is defined by the summation of the orders of vertices, namely ghost numbers of coherent open string fields, as \begin{equation}l([\mathcal{F}_{*_1*_2}])=\sum_{V\in\partial[\mathcal{F}_{*_1*_2}]}\mathrm{ord}(V)\ \ge0.\end{equation}

 The $*$ product $[\mathcal{F}_{ij}^{l_1}]*[\mathcal{F}_{jk}^{l_2}]$ is a formal summation of the classes of flows with the values of the KP/Toda $\tau$ function:\begin{equation}[\mathcal{F}_{ij}^{l_1}]*[\mathcal{F}_{jk}^{l_2}]:=\sum_{\mathrm{ord}(V^p)\ge0}(-)^{l_3}\cdot\tau(\mathcal{F}^g(V^p))[\mathcal{F}_{ik}^{p,\ l_3}]\end{equation}
where ${l_3:=l_1+l_2-2\sum_{\partial[\mathcal{F}_{ij}^{l_1}]\cap\partial[\mathcal{F}_{jk}^{l_2}]}\mathrm{ord}(V)}$. 

  We define the boundary operator $\delta$ of the (co)chain complex $(\mathcal{C}_*^C,\delta)$, i.e., 
 \begin{equation}
 \delta:\mathcal{C}_*^C(c_i,c_j)\to \mathcal{C}_{*\mp1}^C(c_i,c_j),\ \delta^2=0
 \end{equation} as 
  \begin{equation}
  \delta[\mathcal{F}_{ij}^p]:=\sum_{V}(-)^{p\mp1}\cdot\tau([\mathcal{F}^p]*[\mathcal{F}^{p\mp1}]^{-1})[\mathcal{F}_{ij}^{p\mp1}].
  \end{equation}
Since the identity for the partition function $Z$ \begin{equation}{\sum_{V_1,V_2}\mathrm{sgn}\cdot Z([\mathcal{F}^{*_1}])\times Z([\mathcal{F}^{*_2}]^{-1})|_{\mathrm{ord}(V_1-V_2)=*_1-*_2}=\sum_{V}\mathrm{sgn}\cdot Z([\mathcal{F}^{*_1}]*[\mathcal{F}^{*_2}]^{-1})}\end{equation} stands, we obtain the non-trivial boundary condition of $\delta$: 
\begin{eqnarray}
\langle\delta^2[\mathcal{F}^p],[\mathcal{F}^{p\mp2}]\rangle&&=\sum_{l([\mathcal{F}^{p\mp1}])=l([\mathcal{F}^{p}])\mp1}Z[[\mathcal{F}^p]*[\mathcal{F}^{p\mp1}]^{-1}]\times Z[[\mathcal{F}^{p\mp1}]*[\mathcal{F}^{p\mp2}]^{-1}]\nonumber\\&&=\sum_{\partial[\mathcal{F}^{p}]\cap\partial[\mathcal{F}^{p\mp1}]^{-1}}Z[[\mathcal{F}^p]*[\mathcal{F}^{p\mp1}]^{-1}]\times Z[[\mathcal{F}^{p\mp1}]*[\mathcal{F}^{p\mp2}]^{-1}]\nonumber\\&&+\sum_{\partial[\mathcal{F}^{p\mp1}]\cap\partial[\mathcal{F}^{p\mp2}]^{-1}}Z[[\mathcal{F}^p]*[\mathcal{F}^{p\mp1}]^{-1}]\times Z[[\mathcal{F}^{p\mp1}]*[\mathcal{F}^{p\mp2}]^{-1}]\nonumber\\&&=((-)^{p}+(-)^{p\mp1})\cdot Z[[\mathcal{F}^p]*[\mathcal{F}^{p\mp2}]^{-1}]=0.
\end{eqnarray}
  
In addition, we define the Massey product $\eta_k$ with the coefficient $\tau_k$ on the (co)chain complex $\mathcal{C}^C_*$, where the summation is taken over all orders of the $*$ product:
\begin{eqnarray}
&&\eta_k([\mathcal{F}_{01}^{l_1}]\otimes\cdots\otimes[\mathcal{F}_{k-1\ k}^{l_k}])\nonumber\\&&:=\sum_{[\mathcal{F}_{k\ 0}^{l_{k+1}}]=[\mathcal{F}_{01}^{l_1}]*[\mathcal{F}_{12}^{l_2}]*\cdots*[\mathcal{F}_{k-1\ k}^{l_k}]}\sum_{V}\mathrm{sgn}([\mathcal{F}_{k\ 0}^{l_{k+1}}])\cdot\tau_k(\mathcal{F}^g(V))[\mathcal{F}_{k\ 0}^{l_{k+1}}].
\end{eqnarray}

We check the $A_{\infty}$ associativity of this category.

The WDVV equation concerning third derivations of free energy
\begin{equation}\mathcal{F}_{ikm}\eta^{kl}\mathcal{F}_{ljn}=\mathcal{F}_{ijk}\eta^{kl}\mathcal{F}_{lmn}\end{equation} mathematically implies the incomplete associative condition of the $*$ product, whose incompleteness originates in the extra moduli parameter, i.e., the summation over $\eta^{*_1*_2}$ in eq.(32) which physically corresponds to the Feynman integral ${\int dp\frac{1}{p^{*_1}p_{*_2}\eta_{*_1}^{*_2}-m^2}}$ and reflects the first term in eq.(33):
\begin{equation}(\partial\eta_3)([\mathcal{F}_{ij}^{l_1}]\otimes[\mathcal{F}_{jk}^{l_2}]\otimes[\mathcal{F}_{kl}^{l_3}])\pm([\mathcal{F}_{ij}^{l_1}]*[\mathcal{F}_{jk}^{l_2}])*[\mathcal{F}_{kl}^{l_3}]\pm[\mathcal{F}_{ij}^{l_1}]*([\mathcal{F}_{jk}^{l_2}]*[\mathcal{F}_{kl}^{l_3}])=0.
\end{equation}
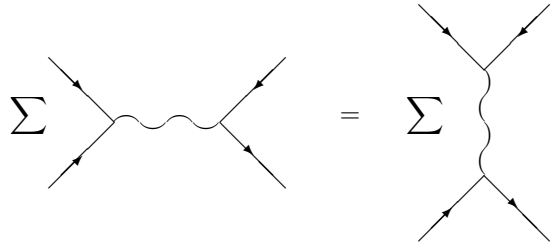
\begin{figure}[ph]
\begin{center}
\begin{picture}(10000,6500)(0,0)
\put(-8000,3000){$\displaystyle{\sum}$}
\put(4500,3000){$=$}
\put(7000,3000){$\displaystyle{\sum}$}
\drawline\photon[\E\REG](-4000,3000)[4]

\drawline\fermion[\SE\REG](\particlebackx,\particlebacky)[3500]
\drawarrow[\SE\ATTIP](\pmidx,\pmidy)
\drawline\fermion[\NE\REG](\photonbackx,\photonbacky)[3500]
\drawarrow[\SW\ATTIP](\pmidx,\pmidy)
\drawline\fermion[\NW\REG](\photonfrontx,\photonfronty)[3500]
\drawarrow[\SE\ATTIP](\pmidx,\pmidy)
\drawline\fermion[\SW\REG](\photonfrontx,\photonfronty)[3500]
\drawarrow[\NE\ATTIP](\pmidx,\pmidy)

\drawline\photon[\S\REG](10000,5000)[4]

\drawline\fermion[\SE\REG](\particlebackx,\particlebacky)[3500]
\drawarrow[\SE\ATTIP](\pmidx,\pmidy)
\drawline\fermion[\SW\REG](\photonbackx,\photonbacky)[3500]
\drawarrow[\NE\ATTIP](\pmidx,\pmidy)
\drawline\fermion[\NW\REG](\photonfrontx,\photonfronty)[3500]
\drawarrow[\SE\ATTIP](\pmidx,\pmidy)
\drawline\fermion[\NE\REG](\photonfrontx,\photonfronty)[3500]
\drawarrow[\SW\ATTIP](\pmidx,\pmidy)
\end{picture}
\end{center}
\caption{Schematic drawing of the WDVV equation without 2-loops}
\end{figure}
This relation is the first $A_{\infty}$ condition eq.(4).

The incompleteness of eq.(33) is expanded to the higher associativity in the perturbation of the partition function, i.e, the second $A_{\infty}$ condition eq.(5):
\begin{eqnarray}&&(\partial\eta_k)([\mathcal{F}_{01}^{l_1}]\otimes\cdots\otimes [\mathcal{F}_{k-1\, k}^{l_k}])\nonumber\\
&&=\sum_{1\le i\le j\le k}\pm\eta_{k-j+i}([\mathcal{F}_{01}^{l_1}]\otimes \cdots\otimes\eta_{j-i}([\mathcal{F}_{i-1\, i}^{l_i}]\otimes \cdots\otimes [\mathcal{F}_{j-1\, j}^{l_j}])\otimes\cdots\otimes [\mathcal{F}_{k-1\, k}^{l_k}]).
\end{eqnarray}

With the above formulae, we have completed the main statement.
 \section{Planar Homological Mirror Symmetry}
In this section, we formularize planar limited homological mirror symmetry based on the Dijkgraaf-Vafa conjecture.
\subsection{Brief review of Dijkgraaf-Vafa theory}
In this section, we review the Dijkgraaf-Vafa conjecture for the {B}-model.

\smallskip
\smallskip
\textit{
The Dijkgraaf-Vafa conjecture states that the action of ${B}$-model topological open string fields as holomorphic Chern-Simons fields on a Calabi-Yau manifold under the conditions of geometric transition\cite{A,A1} is equal to that of a one matrix model with a potential $W$, $S=\frac{1}{g_s}TrW(\Phi)$.
}
\footnote{In this subsection, we omit consideration of the coupling constant.}
\smallskip
\smallskip

First, we explain the Dijkgraaf-Vafa duality that exists between a $N\times N$ matrix model and Seiberg-Witten geometry obtained from $\mathcal{N}=2$ $U(N)$ super Yang-Mills theory.

We consider $\mathcal{N}=2$ $U(N)$ super Yang-Mills theory with a tree level superpotential $W$. This becomes $\mathcal{N}=1$ super Yang-Mills theory, but under the condition
\begin{equation}
\mathrm{deg} W=N+1,\ W\to\epsilon W,\ \epsilon \to0,
\end{equation} the broken $\mathcal{N}=2$ symmetry recovers. Under this assumption (35), the rank of the matrix variable exceptionally matches the degree of potential $W$. \cite{N,N2} In the following discussion, we consistently assume the condition (35).

According to Refs. 1,2 and 4-6, we construct Seiberg-Witten data. Here, we consider a one matrix model which consists of the matrix $\Phi$ as a matrix model variable.

 For a $N\times N$ Hermitian matrix variable denoted by $\Phi$ in one matrix model, the matrix partition function for the action $TrW(\Phi)$ is defined as
\begin{equation}Z\!=\!\int\!d{\Phi}\exp(-TrW(\Phi)).
\end{equation}
This partition function $Z$ is rewritten using eigenvalues $\lambda_s$ of the matrix $\Phi$ as
\begin{equation}Z\!\sim \!\int\!\prod_sd\lambda_s\exp(-W(\lambda)+\sum_{i,j}\log(\lambda_i-\lambda_j)),
\end{equation}
and the eigenvalue density $\rho(\lambda)d\lambda=\sum_s\delta(\lambda-\lambda_s)d\lambda$ as
\begin{equation}Z\!\sim\!\int\!\mathscr{D}\rho(\lambda)\exp(-\int\rho(\lambda)W(\lambda)\log(\lambda-\lambda\prime)d\lambda d\lambda\prime).
\end{equation}
The classical equation of motion of this action is a differential equation of $\rho(\lambda)$:
\begin{equation}
W\prime(\lambda)=\oint\frac{\rho(\lambda\prime)}{\lambda-\lambda\prime},
\end{equation}
and becomes an algebraic equation at the large $N$ limit. Its solution $\rho_0(\lambda)$ is interpreted as a Riemann surface.
More concretely, we set the function $y(\lambda)$ as
\begin{equation}
y(\lambda)=\sum_i\frac{1}{\lambda_i-\lambda}+W\prime(\lambda)\!\sim\!2\int\frac{\rho(\lambda)}{\lambda\prime-\lambda}d\lambda\prime+W\prime(\lambda).
\end{equation}
By defining the $N-1$-degree polynomial $f_{N-1}=\int\frac{W\prime(\lambda)-W\prime(\lambda\prime)}{\lambda-\lambda\prime}\rho(\lambda)d\lambda\prime$, the spectral curve $C$ becomes
\begin{equation}
C:y^2(\lambda)=W\prime^2(\lambda)+f_{N-1}(\lambda).
\end{equation}
This curve $C$ has $N$ branch cuts and two kinds of period.

We set the Seiberg-Witten moduli space as the moduli space of the solution of the Seiberg-Witten curve $C$:
\begin{equation}
\mathcal{M}\!=\!Set\{y(\lambda)\}d\lambda,
\end{equation}
 the Seiberg-Witten differential has a $1$-form: 
\begin{equation}
dS\!=\!y(\lambda)d\lambda,
\end{equation}
and we set the prepotential as the matrix free energy.

In the above matrix integral eq.(37), the $U(N)$ gauge group breaks to the $U(1)^N$ group, but classical mechanically it remains the $U(N)$ group. Therefore, if we add the $U(1)$ fixing condition, the number of Seiberg-Witten moduli variables becomes $N-1$. 

On the other hand, by symmetry breaking $U(N)\to SU(N)$, we add the condition ${\prod_{i}\lambda_i=1}$. Thus, the 
spectral curve $C$ becomes a Riemann surface with genus $N-1$ as discussed in sec.2.3. We note that the number of degrees of freedom of the eigenvalues of the matrix is also reduced to $N-1$.
This fact is consistent with the result obtained from the Seiberg-Witten curve. This consistency originates in the condition of the geometric transition.\cite{A,A1,N,N2}

As noted in sec.2.3, $SU(N)$ Seiberg-Witten data can be captured in the context of a Whitham deformation of an integrable system, i.e., KP/Toda lattice hierarchy.
Thus, from the Dijkgraaf-Vafa conjecture, we can infer that the Massey product in the $A_{\infty}$ category $\mathcal{D}_{\infty}^b(Coh(V^{\dagger}))$ matches one of $\mathcal{D}_{\infty}\mathcal{F}$ as a partition function, and under Dijkgraaf-Vafa conditions, the number of degrees of freedom of the D-branes matches one of the Seiberg-Witten moduli because of the reason given above.
There are natural correspondences between BPS monopole states and their deformations in moduli space, and {B}-branes and their Chan-Paton factors, i.e., open strings.
This evidence leads to this article's formulation.
\subsection{Equivalence to large-$N$ reduced matrix model}
Here, to clarify the relation between matrix model and open string fields theory, we detail the background of the above Dijkgraaf-Vafa conjecture as a large-$N$ reduced matrix model. \cite{B,O,P} 
 We review large $N$ reduced matrix model description of Dijkgraaf-Vafa duality according to Ref. 3.

In this section, we need the action of noncommutative $U(N)$ gauge theory for the gauge potential $\mathscr{A}^{\mu}$ also denoted by $\mathscr{M}^{\mu}$ coupled to fermion $\Psi$ in the adjoint representation: \cite{B}
\begin{equation}
S=\frac{1}{4g_{YM}^2}Tr(\frac{1}{4}[\mathscr{A}^{\mu},\mathscr{A}^{\nu}]^2+\frac{1}{2}\bar{\Psi}\Gamma^{\mu}[\mathscr{A}^{\mu},\Psi]).
\end{equation}

As proved in Ref. 3, this action is able to be extended to the action of a supermatrix model with chiral superfields $\Phi$ as Chan-Paton factors on D-branes in a large-$N$ reduced model described by the supertrace of supermatrix $Str_{\alpha}(\Phi)=2Tr(\sigma^3\Phi)$, for noncommutativity of coordinates $\alpha=x^{\mu},\theta^{\mu}$ on noncommutative space which satisfy $[\alpha^1,\alpha^2]=2i\gamma\sigma^3$, and superpotential $W$ as
\begin{equation}
S^{hol}=Str_{x\otimes\theta}W(\Phi).
\end{equation}

The notation $hol$ denotes that this action $S^{hol}$ is a holomorphic term of one of a background \text{super} large-$N$ (twisted) reduced model composed of matrices $\Phi$, $\bar{\Phi}$, $\mathscr{V}$, $W^{\alpha}$. These components correspond to adjoint matter superchiral (anti-)fields, gauge fields and field strengths. \cite{B}

This action $S^{hol}$ is equivalent to the action of the original Dijkgraaf-Vafa theory of an ordinary one matrix model, $Tr(W(\Phi))$.

Next, we explain the correspondence between the D-branes system and a matrix variable in a large-$N$ reduced matrix model. Essentially, the situation does not change in a super large-$N$ reduced matrix model.

In the bosonic matrix $\mathscr{M}^{\mu}$, coordinates of D-branes are diagonal elements $\mathscr{D}^{\mu}_{nn}$ of matrix $\mathscr{M}^{\mu}$. Then, nondiagonal elements $s^{\mu}_{mn}$ of $\mathscr{M}^{\mu}$ are open strings running from D-brane $\mathscr{D}^{\mu}_{mm}$ to D-brane $\mathscr{D}^{\mu}_{nn}$.

According to Ref. 31, this latter conclusion is justified by the following.

We consider an Hermitian matrix $\mathscr{M}^{\mu}$ of a field perpendicular to a D-brane in this matrix model.
If $s_{ij}^m$ is a variable of a string field, its mass is proportional to the length between D-branes $i$ and $j$. In field theory, this mass is proportional to a coefficient of $(\mathscr{D}^{\mu}_{ij})^2$ in the action. In the above action, the potential term $Tr[\mathscr{M}^{\mu},\mathscr{M}^{\mu}]^2$ is proportional to $(\bar{\mathscr{D}}^{\mu}_i-\bar{\mathscr{D}}^{\mu}_j)^2$, for the vacuum expectancy $\bar{\mathscr{M}}_i^{\mu}$.

From the above, we define the Dijkgraaf-Vafa reduction $\mathcal{T}$.

Firstly, this does not change the KP/Toda $\tau$ function:
\begin{equation}\mathcal{T}:\tau\to\tau.\end{equation}

Secondly, this transforms the alteration domains of the KP/Toda $\tau$ function from the Hilbert space of a one matrix model through the transformation of its spectral sets $\Lambda$ to another alteration domain, the second quantized Hilbert space of the Toda lattice integrable system, $\mathscr{H}_{Toda}$ \cite{R}
\begin{equation}\mathcal{T}:\mathscr{H}_{\Phi}\!\to \mathscr{H}_{Toda}.
\end{equation}

 We shall define the elements of Hilbert space $\mathscr{H}_{Toda}$ in sec.3.3 as the Hilbert space $\mathscr{H}_{\Phi}$ of one matrix variable solutions of the covariant equation.

By summarizing the first statement in the introduction and $A_{\infty}$-enhanced homological mirror symmetry $\mathcal{D}_{\infty}Fuk(V)\simeq \mathcal{D}_{\infty}^b(Coh(V^{\dagger}))$,\cite{E} we formularize topological open string field theory as an algebraic equation containing two types of higher composition matrix in $A_{\infty}$-enhanced-derived $A_{\infty}$ categories $\mathcal{D}_{\infty}Fuk(V)$ and $\mathcal{D}_{\infty}\mathcal{F}(C)$, as mentioned in sec.2.1. 
\subsection{Further details of eq.(2)}
The equation of homological mirror symmetry for topological planar open string field theories
satisfies the following three physical conditions.
\begin{enumerate}
\item	It stands for perturbation of any degree $k$.

\item	It stands for any strong/weak coupling constant.
\item It includes gravity/gauge theory duality.
\end{enumerate}
 \begin{equation}\mathfrak{F}(\lambda\,|\>\Phi^{\dagger})\simeq\tau(\mathcal{T}\!\cdot\!\Phi).\end{equation}
We write eq.(48) using $\simeq$ not $=$ since this formula describes the equivalence between two $A_{\infty}$ algebras of higher composition structures which satisfy eq.(4) and eq.(5) in each $A_{\infty}$ category. 

In eq.(48), we let $\Phi^{\dagger}$ be a matrix variable in the Chern-Simons one matrix model proposed in Ref. 33 which describes the mirror contents of $\Phi$, and satisfy the condition $(\Phi^{\dagger})^{\dagger}\simeq \Phi$, where $\Phi \in Her(N)$ is a matrix variable in a one matrix model which satisfies the conditions of the Dijkgraaf-Vafa conjecture. 

We note that in the mirror symmetry involution $\dagger$, eigenvalues are not always conserved. For example, generally the {A}-brane and the corresponding {B}-brane have different dimensions and configurations to each other.\cite{F,F1}

It is shown in Ref. 33 that the partition function of the Chern-Simons one matrix model with the action $TrW(\Phi^{\dagger})$ matches the partition function of the topological open string {A}-model:
\begin{equation}
Z_{{A}}^{inst}\!=\!\int\!d{\Phi^{\dagger}}\exp(-TrW(\Phi^{\dagger})).
\end{equation}
Furthermore, in the situation of the Chern-Simons one matrix model, the $n$-th eigenvalue $\lambda_n$ of the matrix variable $\Phi^{\dagger}$ describes the measure of the $n$-th {A}-brane, i.e., the variable of the $n$-th sphere on the mirror manifold of resolved conifold.\cite{A,A1,T} Thus, in eq. (48), we let $\mathfrak{F}(\lambda\,|\>\Phi^{\dagger})$ denote the partition function $Z_{{A}}$.

We can easily check that eq.(48) satisfies the above three conditions.

Condition (\roman{ichi}) is realized by the $A_{\infty}$ relations of eq. (4) and eq. (5) on the partition functions of the $A_{\infty}$ categories $Fuk(V)$ and $\mathcal{F}(C)$.

Condition (\roman{ni}) is realized by $S$-duality on the right-hand side of eq. (48) as the modularity of $\tau$ functions from $Sp(2g,{Z})$ symmetries of Seiberg-Witten curves, and on the left-hand side of eq. (48) as mirror symmetry between the {A}-model and {B}-model in untwisted 
type \Roman{ni}B string theory which reduces the left-hand side to the right-hand side. \cite{T}

Equation. (48) is a multi-linear equation describing the sector of open string fields from the definition of product structures of $A_{\infty}$ algebras.

\section{Gravity/Gauge Theory Duality as Conclusion}
To conclude, we review the contents of this article from the view point of gravity/gauge theory duality.
As seen in the Gopakumar-Vafa duality on deformed / resolved conifolds,\cite{GV} in the planar limit, the open string world sheet becomes the closed string decoupled world sheet. In this article, we formulate the {B}-model side of this phenomenon.
Homological mirror symmetry is originally defined on a D-branes system attached by open strings. The planar limited version implies the mirror symmetry of decoupled closed strings and we formulate this version by using homotopy associative algebra. Then, categorical formulation should satisfy following conditions.
\begin{enumerate}
\item As objects, D-branes are reread as symplectic bases on $N\to\infty$, $SU(N)$ Seiberg-Witten type curves.
\item As morphisms, closed string fields are related to open string fields by the large $N$ duality.
\item The product structures satisfy $A_{\infty}$ algebras that originate in topological gravity.
\end{enumerate}
The physical background geometric transition in the Dijkgraaf-Vafa conjecture satisfies condition $(iii)$. \cite{A,A1,U,U1} 
In this article we retained the mathematical formulation {A}-model side on the mirror manifold of resolved conifold, but, physically this is not same as but rather analogous to the Gopakumar-Vafa duality. 
Therefore, the gravity/gauge theory duality is considered to be fundamental to solving planar homological mirror symmetry.

\bigskip

\leftline{\textit{Acknowledgments.}}

The author wishes to thank Professor Toshiya Kawai for valuable discussions and instruction, Professor Tohru Eguchi for valuable comments, Professor Shu-ichiro Inutsuka and Professor Hikaru Kawai for their kindness, and especially Yoshitaka Yamamoto for continuous encouragement and kindness throughout the years spent preparing this article.
\begin{appendix}
\section{Physical Aspects of Open String Field $A_{\infty}$ Algebra}
In this appendix, we refer to the physical aspects of $A_{\infty}$ structure of open string fields theory.

$A_{\infty}$ algebra of open string fields $A$ without boundary conditions is given by the following higher composition structure.\cite{Nakatsu}

For Hilbert space of cubic open string field $\mathscr{H}$, the higher composition structure $\eta_k$ is given for BRST charge $Q$ and reflector operator $|R_{ab}\rangle\in\mathscr{H}^{\otimes2}$ with indices of two open string disks connecting points $a$ and $b$ that satisfy $\langle A_a|R_{ab}\rangle=|A_b\rangle$ and $k$-vertex $|A_1\ A_2\ \cdots\ A_k\rangle$ geometrical disk $D^2$ with marked $n$ points in this order on the boundary as follows,
\begin{equation}\eta_1(A):=QA,\end{equation}
\begin{equation}\eta_k(A_1,A_2,\cdots,A_k):=(-)^{\eta}\langle A_a\ A_1\ A_2\cdots A_k|R_{ab}\rangle|A_1\rangle|A_2\rangle\cdots|A_k\rangle.\end{equation}
Then, $(\mathscr{H},\eta_k)$ forms an $A_{\infty}$ algebra.

The sketch of the proof of the associativity is as follows. First we apply the BRST charge $Q$ to $\eta_k(A_1,A_2,\cdots,A_k)$ and from the Leibniz rule, 
there emerges two terms such that $Q$ acts to out going $k$-vertex or in coming open string fields. We calculate these terms using the cyclic rule for $k$-vertex
\begin{equation}|A_1\ A_2\ \cdots\ A_k\rangle=(-)^{k+1}|A_2\ A_3\ \cdots\ A_k\ A_1\rangle.\end{equation}
 The $0$-loop open string world sheet is a disk having all perturbation degrees, and thus has these cyclic constraints. 
As a result of the calculation detailed in Ref. 38, the $k$-th higher composition of open string fields homotopy associative algebra is written by degree $k$ disk amplitudes.

Finally, we comment on the above representation of product structure which becomes the KdV $\tau$ function on the resolved conifold using the description of {B}-model coherent open string fields by bosonic fields as detailed in Ref. 39.
This fact suggests the main statement truly describes open string field theory on a resolved conifold. 
\end{appendix}

\end{document}